\documentclass[aps,prc,twocolumn,groupedaddress,showpacs,10pt,dvipsnames]{revtex4-1}

\usepackage[utf8]{inputenc}
\usepackage{amsfonts}
\usepackage{amsmath}
\usepackage{amssymb}
\usepackage[pdftex]{graphicx}
\usepackage{xcolor}
\usepackage{tikz}
\usepackage{cancel}
\usetikzlibrary{arrows,patterns}

\begin{document}

\newcommand{\tj}[6]{ \begin{pmatrix}
  #1 & #2 & #3 \\
  #4 & #5 & #6
 \end{pmatrix}}


\title{$^9$Be elastic scattering on $^{208}\text{Pb}$ and $^{27}\text{Al}$ within a four-body reaction framework} 




\author{J. Casal}
\email{jcasal@us.es}
\author{M. Rodr\'{\i}guez-Gallardo}
\author{J. M. Arias}
\affiliation{Departamento de F\'{\i}sica At\'omica, Molecular y Nuclear,
  Facultad de F\'{\i}sica, Universidad de Sevilla, Apartado 1065, E-41080
  Sevilla, Spain} 


\date{\today}

\begin{abstract}
We investigate the low-energy $^9$Be elastic scattering on two different targets (heavy, light) within a four-body framework using the Continuum-Discretized Coupled-Channels (CDCC) method. The $^9$Be projectile is described in a $\alpha + \alpha + n$ three-body model using the analytical transformed harmonic oscillator (THO) basis in hyperspherical coordinates. We show that continuum couplings are important to describe the elastic cross section, especially at low energies and on heavy targets. The dipolar contribution to the elastic cross section at energies around the Coulomb barrier is important but small compared to the case of halo nuclei. The effect of the projectile low-energy resonances is also relevant.
The agreement with the available experimental data supports the reliability of the method to describe reactions induced by three-body projectiles including more than one charged particle.

\end{abstract}


\maketitle


  \section{Introduction}
  
  Over the last decades, the Continuum-Discretized Coupled-Channels (CDCC) method~\cite{Yahiro86,Austern87} has been applied as a fundamental tool to study nuclear reactions involving weakly-bound nuclei. In particular, the method 
  has been very useful in the description of elastic and breakup observables for reactions involving halo nuclei~\cite{Hansen95}. Loosely bound systems are easily broken up in the scattering process due to the nuclear and 
  Coulomb forces. This effect can be properly treated within the CDCC formalism including the coupling to the continuum part of the spectrum or breakup channels~\cite{Matsumoto06,MRoGa08,JALay10,Juanpi13}.
  
  For two-body projectiles, whose continuum states can be easily calculated, the traditional discretization method is the \textsl{binning} procedure, in which the continuum spectrum is truncated at a maximum excitation energy and divided into energy (or momentum) intervals. For each  interval, or bin, a normalizable state is built up by superposition of the scattering states within the interval. This method has been extended to three-body projectiles such as 
  the halo nuclei $^6$He~\cite{MRoGa09} and $^{11}$Li~\cite{Juanpi13}, which consist of a core and two valence neutrons. For three-body systems with more than one charged particle, however, the calculation of 
  continuum states is a very involved problem~\cite{Nguyen2} since the asymptotic behavior of the wave functions is not known in general. An alternative to the binning procedure is the so called Pseudo-State (PS) method, 
  which consists in representing the continuum spectrum of the projectile by the eigenstates of its internal Hamiltonian in a basis of square-integrable functions. The advantage of this procedure is that it does not require 
  going through the exact continuum wave functions, and the knowledge of the asymptotic behavior is not needed. A variety of bases has been proposed for two-body~\cite{HaziTaylor70,Matsumoto03,MRoGa04,AMoro09} and also for 
  three-body~\cite{Desc03,Matsumoto04,MRoGa05,JCasal13} calculations. The PS discretization has been successfully applied to CDCC calculations involving three-body projectiles~\cite{Matsumoto06,MRoGa08} and, more recently, to the description of reactions 
induced by two-body projectiles with core excitations~\cite{RdDiego14XCDCC}.
  
  In this work, we investigate $^9$Be scattering on two different targets using the CDCC formalism. In a three-body model ($\alpha+\alpha+n$) this nucleus shows a Borromean structure, since none of its binary subsystems, 
  $\alpha+n$ or $\alpha + \alpha$, is bound. Two-body models for $^9$Be need to assume a $^8$Be + $n$ or $^5$He + $\alpha$ cluster structure, while both configurations are naturally included in a consistent three-body model. 
  For that reason, we use the four-body CDCC framework considering three-body projectile internal states. The $^9$Be system is stable but has a small separation energy~\cite{Tilley04}, so breakup effects are expected to be 
  important for the description of reactions induced by this nucleus. This has already been reported in previous CDCC calculations using a two-body approximation for $^9$Be~\cite{Pandit11} and, more recently, with a 
  three-body model for the scattering on a $^{208}$Pb target~\cite{Descouvemont1,Descouvemont15}.
  The elastic and breakup data on $^9$Be + $^{208}$Pb~\cite{Wolliscroft04,Yu10} and $^{27}$Al~\cite{Gomes04} provide a good opportunity to test the three-body description of the $^9$Be system within a four-body 
  reaction formalism and to study the effect of different target masses on the reaction mechanism.
  
  We describe the $^9$Be three-body system using the PS discretization method. In a recent work~\cite{JCasalAIP12,JCasal13} we presented a PS method for three-body systems based on an analytical local scale 
  transformation of the harmonic oscillator (HO) basis, the transformed harmonic oscillator (THO) method. We successfully applied this procedure to $^6$He ($\alpha + n + n$) and, more recently, to 
  $^9$Be ($\alpha+\alpha+n$)~\cite{JCasal14}, focusing on their astrophysical interest. The analytical THO basis provided a reliable description of the $^9$Be properties, which encourages its application to reactions induced 
  by this nucleus. 
  
  This paper is structured as follows. In Sec.~\ref{sec:theory} the analytical THO method for three-body systems is briefly presented, and the main features of the four-body CDCC method are shown. In Sec.~\ref{sec:application} 
  the formalism is applied to the particular case of $^9$Be, focusing on the $^9$Be + $^{208}$Pb and $^9$Be + $^{27}$Al elastic scattering at different energies. Finally, in Sec.~\ref{sec:summary}, the main conclusions of this work are summarized.

\section{Theoretical formalism}\label{sec:theory}

\begin{figure}
\centering

 \includegraphics[width=\linewidth]{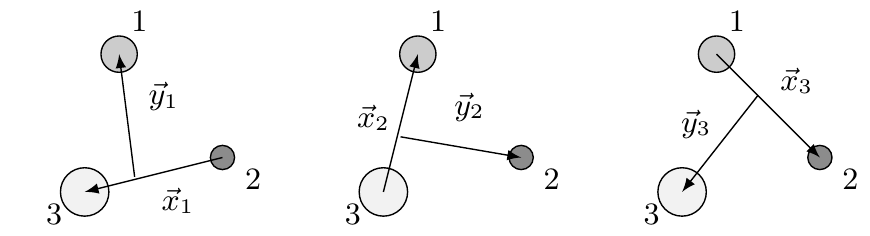}
 \caption{The three sets of scaled Jacobi coordinates.}
 \label{fig:sets}
\end{figure}
The three- and four-body formalisms used in this work are described in detail in Refs.~\cite{JCasal13,JCasal14,MRoGa08}. For completeness, in this section we summarize the main features and establish the notation we use throughout the paper. 

We describe the three-body projectile using Jacobi coordinates $\{\boldsymbol{x}_k,\boldsymbol{y}_k\}$. The variable $\boldsymbol{x}_k$ is proportional to the relative coordinate between two particles, and $\boldsymbol{y}_k$ 
is proportional to the distance from the center of mass of the $x$-subsystem to the third particle, both with a scaling factor depending on their masses~\cite{MRoGa05}. Label $k$ identifies one of the three possible Jacobi 
systems, as shown in Fig.~\ref{fig:sets}. As in Ref.~\cite{JCasal14}, we use the notation in which, for example, the Jacobi-1 system corresponds to the system where the particles (2,3) are related by the coordinate 
$\boldsymbol{x}_1$. From the Jacobi coordinates we can define the hyperspherical coordinates $\{\rho,\alpha_k,\widehat{x}_k,\widehat{y}_k\}$, where the hyper-radius $(\rho)$ and the hyperangle $(\alpha_k)$ are given by
\begin{align}
\rho = & \sqrt{x_k^2 + y_k^2}, \\
\alpha_k = & \tan\left(\frac{x_k}{y_k}\right),
\label{eq:jachyp}
\end{align}
and $\{\widehat{x}_k,\widehat{y}_k\}$ are the two-dimensional angular variables associated to $\{\boldsymbol{x}_k,\boldsymbol{y}_k\}$. Note that, while the hyperangle depends on $k$, the hyper-radius is the same for the three 
Jacobi systems.

\subsection{Analytical THO method for three-body systems}

PS methods consist in diagonalizing the Hamiltonian in a discrete basis of square-integrable functions. Using hyperspherical coordinates, the solutions of the three-body Schrödinger equation in one of the Jacobi systems can 
be expanded as (if $k$ is fixed we do not specify it)
\begin{equation}
 \phi_{nj\mu}(\rho,\Omega)=\rho^{-5/2}\sum_{\beta} \chi_{n\beta}^{j\mu}(\rho)\mathcal{Y}_{\beta j\mu}(\Omega),
 \label{eq:wf}
\end{equation}
where the label $n$ enumerates the eigenstates, $\Omega\equiv\{\alpha,\widehat{x},\widehat{y}\}$ is introduced for the angular dependence, and $\beta\equiv\{K,l_x,l_y,l,S_x,j_{ab}\}$ is a set of quantum numbers called channel. 
In this set, $K$ is the hypermomentum, $l_x$ and $l_y$ are the orbital angular momenta associated with the Jacobi coordinates $\boldsymbol{x}$ and $\boldsymbol{y}$, respectively, $l$  is the total orbital angular momentum 
($\boldsymbol{l}=\boldsymbol{l_x}+\boldsymbol{l_y}$), $S_x$ is the spin of the particles related by the coordinate $\boldsymbol{x}$, and $j_{ab}$ results from the coupling 
$\boldsymbol{j_{ab}}=\boldsymbol{l}+\boldsymbol{S_x}$. If we denote by $I$  the spin of the third particle, that we assume to be fixed, the total angular momentum is $\boldsymbol{j}=\boldsymbol{j_{ab}} + \boldsymbol{I}$.
The functions $\mathcal{Y}_{\beta j\mu}(\Omega)$ are states of good total angular momentum, expanded in hyperspherical harmonics (HH)~\cite{Zhukov93}. See, for instance, Appendix A in Ref.~\cite{JCasal14}.

We expand the hyperradial functions $\chi_{n\beta}^{j\mu}(\rho)$ in the analytical THO basis as
\begin{equation}
 \chi_{n\beta}^{j\mu}(\rho) = \sum_{i=0}^{i_{\rm max}} C_n^{i\beta j} U_{i\beta}^\text{THO}(\rho),
 \label{eq:expandTHO}
\end{equation}
where $i$ denotes the hyperradial excitation, so that $(i_{\rm max}+1)$ represents the number of hyperradial functions included for each channel. The THO basis functions are based on a local scale transformation, $s(\rho)$, of the HO functions, 
\begin{equation}
  U_{i\beta}^{\text{THO}}(\rho)=\sqrt{\frac{ds}{d\rho}}U_{iK}^{\text{HO}}[s(\rho)].
\label{eq:R}
\end{equation}
The transformation keeps the simplicity of the HO functions, but converts their Gaussian asymptotic behavior into an exponential one. This provides a suitable representation of bound and resonant states to calculate structure and scattering observables. In this paper, as in previous publications~\cite{JCasal13,JCasal13,JALay10,AMoro09}, we use the analytical form proposed by Karataglidis \textsl{et al.}~\cite{Karataglidis},
\begin{equation}
s(\rho) = \frac{1}{\sqrt{2}b}\left[\frac{1}{\left(\frac{1}{\rho}\right)^{\xi} +
\left(\frac{1}{\gamma\sqrt{\rho}}\right)^\xi}\right]^{\frac{1}{\xi}},
\label{eq:LST}
\end{equation}
depending on the parameters $\xi$, $\gamma$ and $b$. We have fixed for all calculations $\xi=4$ as in Ref.~\cite{JCasal14}, since it was found previously a very weak dependence of the results on this parameter. Note that the THO hyperradial wave functions depend, in general, on all the quantum numbers included in a channel $\beta$, however the HO 
hyperradial wave functions only depend on the hypermomentum $K$. The eigenstates are obtained by inserting Eq.~(\ref{eq:expandTHO}) in Eq.~(\ref{eq:wf}),
\begin{equation}
 \phi^{\rm THO}_{nj\mu}(\rho,\Omega)=\rho^{-5/2}\sum_{\beta} \sum_{i=0}^{i_{\rm max}} C_n^{i\beta j} U_{i\beta}^\text{THO}(\rho)\mathcal{Y}_{\beta j\mu}(\Omega),
 \label{eq:wfTHO}
\end{equation}
where $C_n^{i\beta j}$ are just the diagonalization coefficients, and the associated energy eigenvalue is denoted by $\varepsilon_{nj}$. We refer the reader to Ref.~\cite{JCasal13} for details about the Hamiltonian matrix element calculations.

The function $s(\rho)$ behaves asymptotically as $\frac{\gamma}{b}\sqrt{\frac{\rho}{2}}$, and hence the THO hyperradial wave functions obtained behave at large distances as 
$\exp{(-\gamma^2\rho/2b^2)}$. Therefore, the ratio $\gamma/b$ governs the asymptotic behavior of the THO functions: as $\gamma/b$ increases, the hyperradial extension of the basis 
decreases, and some of the eigenvalues obtained by diagonalizing the Hamiltonian explore higher energies~\cite{JALay10}. That is, $\gamma/b$ determines the density of PSs as a function of 
the energy. This allows to select an optimal basis depending on the observable of interest.

\subsection{Four-body CDCC framework}

\begin{figure}
 \includegraphics[width=0.8\linewidth]{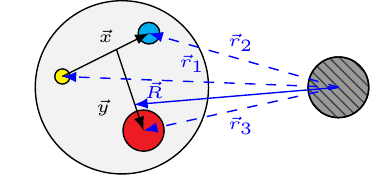}
 \caption{(Color online) Scattering of a three-body projectile by a structureless target.}
 \label{fig:4b}
\end{figure}

We are interested in describing reactions induced by three-body projectiles, illustrated in Fig.~\ref{fig:4b}. The four-body wave-function can be expanded in internal states of the projectile, given by Eq.~(\ref{eq:wfTHO}), as
\begin{equation}
\begin{split}
\Psi_{JM}(\boldsymbol{R},\boldsymbol{x},\boldsymbol{y})  \equiv & \sum_{nj\mu LM_L} \phi_{nj\mu}^{\rm THO}(\boldsymbol{x},\boldsymbol{y})\langle LM_L j\mu|JM\rangle  \\
                                                               & \times i^L Y_{LM_L}(\widehat{R})\frac{1}{R}f_{Lnj}^J(R)
\end{split}
\end{equation}
where $\boldsymbol{R}$ is the coordinate from the target to the center of mass of the projectile, $L$ is the orbital angular momentum of the relative motion, and $J$ is the total angular momentum, given by 
$\boldsymbol{J}=\boldsymbol{L}+\boldsymbol{j}$. The radial functions satisfy then the coupled equations
\begin{equation}
 \begin{split}
  &\left[-\frac{\hbar}{2m_r}\left(\frac{d^2}{dR^2}-\frac{L(L+1)}{R^2}\right)+\varepsilon_{nj}-E\right]f_{Lnj}^J(R) \\
        & + \sum_{L'n'j'}i^{L'-L}V_{Lnj,L'n'j'}^J(R)f_{L'n'j'}^J(R)=0.
  \label{eq:CC}
 \end{split}
\end{equation}
To calculate the coupling potentials $V_{Lnj,L'n'j'}^J(R)$
we use a multipole expansion of the projectile-target interaction $\widehat{V}_{pt}(\boldsymbol{r}_1,\boldsymbol{r}_2,\boldsymbol{r}_3)$. The procedure is 
explained in detail in Ref.~\cite{MRoGa08}. The only difference is that we use here the analytical THO basis to expand the projectile internal states. The parameters of the analytical transformation in Eq.~(\ref{eq:LST}) allow 
us to select an optimal basis to describe the low-energy continuum with a high density of states. This minimizes the convergence problems previously reported for reactions on heavy targets, where the Coulomb repulsion is strong, 
due to the discrete nature of the basis~\cite{MRoGa08}. Besides, the analytical THO basis can be used for three-body projectiles including more than one charged particle, and the previous knowledge of the ground state is not needed.

\section{Application to $^9$Be induced reactions}\label{sec:application}

\subsection{Three-body model for $^9$Be}\label{sec:3b9be}

The $^9$Be nucleus can be described in a three-body model as two alpha particles loosely bound by the remaining neutron. It has a Borromean structure, since its binary subsystems have no bound states. The ground state 
of $^9$Be has total angular momentum $j^\pi=3/2^-$ and a binding energy of 1.5736 MeV below the $\alpha + \alpha + n$ threshold~\cite{Tilley04}. The low-energy spectrum of $^9$Be shows several resonances with angular momenta $j^\pi=1/2^+,5/2^-,1/2^-,5/2^+,3/2^+$ and possibly others,  
whose positions and widths have been investigated by many authors in different experiments~\cite{Sumiyoshi02,Arnold12,Burda10}. The spin-parity assignment for some of these states is still under discussion~\cite{Hirayama15}, although there are no implications for the lowest 1/2$^+$ and 5/2$^-$ resonances. We describe the states of the system using the analytical THO method. The structure 
calculations are reported in Ref.~\cite{JCasal14}, where the method is applied to generate the photodissociation cross section and reaction rate for $^9$Be formation. In these calculations, the $\alpha$-n potential is 
taken from Ref.~\cite{IJThompson00} and the $\alpha$--$\alpha$ potential is the Ali-Bodmer interaction ``a''~\cite{AliBodmer}, modified to reproduce the experimental phase shifts. These are shallow potentials in the sense that they include repulsive terms to remove unphysical two-body  states. Since the three-body calculations are just an approximation to the full many-body problem, including only binary interactions may lead to deviations from the experimental three-body 
energies~\cite{RdDiego10,IJThompson04}. We include then a simple hyperradial three-body force, depending on three parameters, which can be fixed to adjust the position of the known states of the system to the experimental values without distorting 
their structure,
\begin{equation}
 V_{3b}(\rho) = \frac{v_{3b}}{1+\left(\frac{\rho}{r_{3b}}\right)^{a_{3b}}}.
 \label{eq:3bforce}
\end{equation}

Calculations shown in Ref.~\cite{JCasal14} truncate the maximum hypermomentum at large values of $K_{max}$. For $^9$Be, it is necessary to consider $K_{max}\geq 30$ in order to obtain a well 
converged ground state and, more importantly, to achieve converged energy distributions.
For coupled-channels calculations including several $j^\pi$ configurations, however, such $K_{max}$ values  imply to work with very large basis sets, which is computationally challenging.
In this work, we fix smaller $K_{max}$ values and adjust the three-body force parameters to recover the same energy and radius of the relevant states. In the following, 
unless stated otherwise, the calculations presented are performed with $K_{max}=10$. This value provides converged reaction calculations with respect to the hypermomentum, as it will be shown in the following subsection.

\begin{figure}
 \includegraphics[width=1\linewidth]{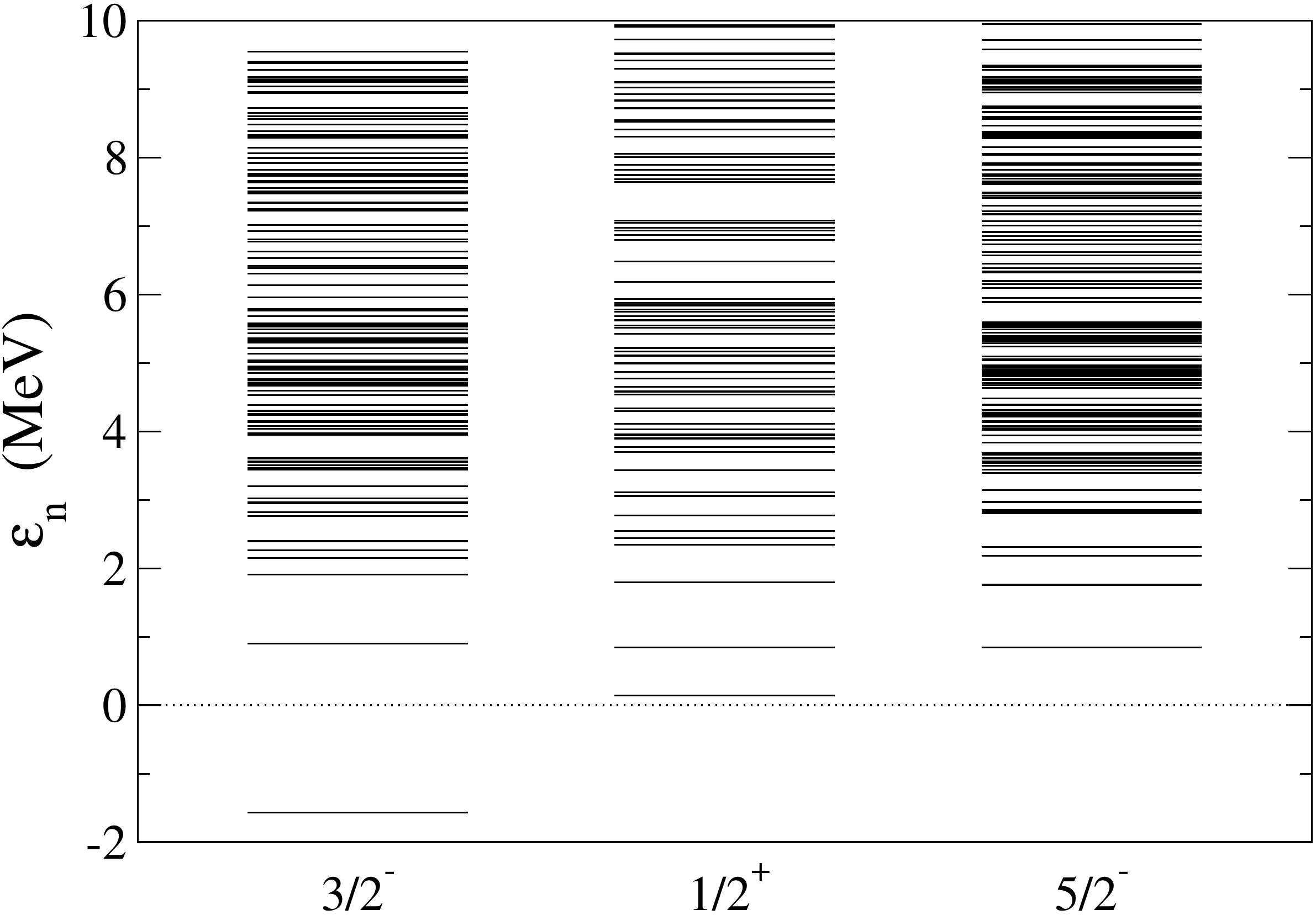}
 \caption{Energy spectra for the states with $j^\pi=3/2^-,1/2^+,5/2^-$ up to 10 MeV. Calculations are performed for an analytical THO basis with $K_{max}=10$ and $i_{max}=8$.}
 \label{fig:spectra}
\end{figure}

With the analytical THO basis, we are able to adjust the parameters of the local scale transformation in order to concentrate more states at low energy. However, if the level density is very high, coupled-channel calculations become more and 
more demanding computationally. We fix the THO parameters to $b=0.7$ fm and $\gamma=1.2$ fm$^{1/2}$ for all the $j^\pi$ states considered. These values ensure a fast convergence of the ground state with respect to the number 
of hyperradial excitations $i_{max}$ and also allows to concentrate a reasonable number of continuum states close to the breakup threshold. As an example, in Fig.~\ref{fig:spectra} we show the energy spectra for different 
$j^\pi$ configurations calculated with $i_{max}=8$. The only negative eigenstate for $j^\pi=3/2^-$ corresponds to the bound state, and the positive energy eigenvalues represent our continuum discretization in the THO basis 
up to 10 MeV.
Assuming that the $\alpha$ particle charge radius is 1.6755 fm, for the ground state we obtain a charge radius of $r_{\rm ch}$ = 2.50 fm. This value is in agreement with the experimental value of 2.519 $\pm$ 0.012 
fm~\cite{Angeli13} and indicates that our description of the system is quite accurate. 
Our model also describes the strong quadrupole deformation of $^9$Be and provides a quadrupole moment of $Q_2=4.82$ e~fm$^2$, close to the experimental value of 5.29 $\pm$ 0.04 e~fm$^2$~\cite{Sundholm91}.

\subsection{$^9$Be $+$ $^{208}$Pb}
In this subsection we study the scattering of $^9$Be on a $^{208}$Pb target at different energies. We use the eigenstates obtained with the analytical THO method to represent the $^9$Be projectile wave functions. States with 
very high excitation energies will not be relevant for the description of the scattering process, since their couplings will be weak. Thus we include in the CDCC calculations only the states up to a given cutoff energy, 
chosen to be high enough to provide converged results.
The data of two different experiments are available in the literature for this reaction~\cite{Wolliscroft04,Yu10}.

The Coulomb barrier for the $^9$Be $+$ $^{208}$Pb system is around 47 MeV. We will show our coupled-channels results above ($E_{lab}=60$ MeV), around (44 MeV) and below (38 MeV) the barrier. Since $^9$Be is a weakly-bound system, we 
expect to observe a strong absorption even at low energies, with an important coupling to breakup channels. This effect is widely known for exotic systems such as halo nuclei~\cite{MRoGa08,Cubero12}. The coupling potentials in 
Eq.~(\ref{eq:CC}) are generated considering the $n$--$^{208}$Pb potential from the Koning and Delaroche global parametrization~\cite{KD} and the $\alpha$--$^{208}$Pb interaction from Ref.~\cite{Barnett74}. Note that these potentials are in general energy-dependent. 
The coupled equations are solved up to $J=301/2$, including projectile-target interaction multipole couplings of the order $Q=0$--5.

\begin{figure}
 \includegraphics[width=1\linewidth]{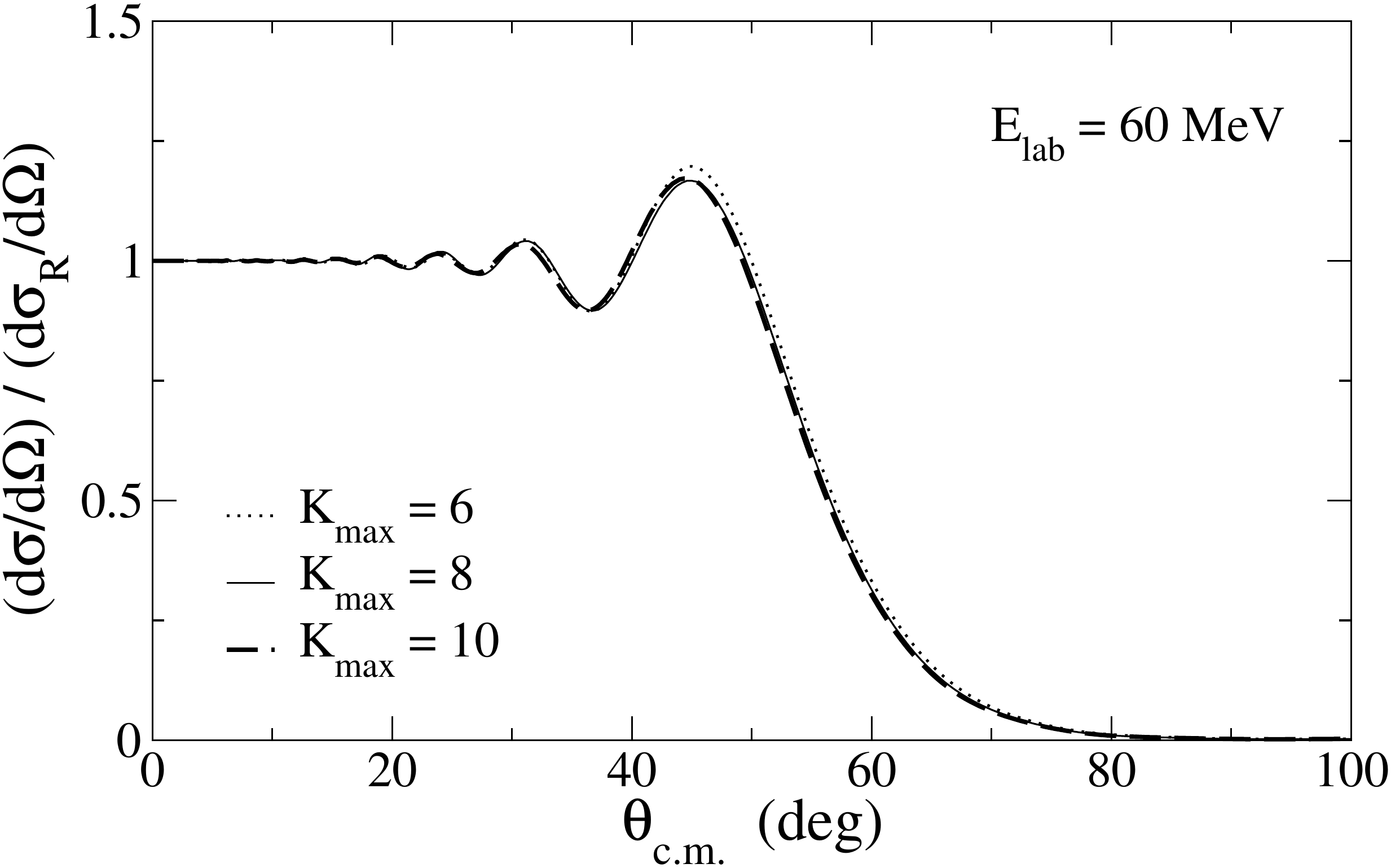}
 \caption{Convergence of the angular distribution of the elastic cross section relative to Rutherford with respect to $K_{max}$ for the reaction $^9$Be + $^{208}$Pb at $E_{lab}=60$ MeV. Calculations are performed with $i_{max}=6$ and $\varepsilon_{max}=6$ MeV.}
 \label{fig:60MeV_kmax}
\end{figure}

\begin{figure}
 \includegraphics[width=1\linewidth]{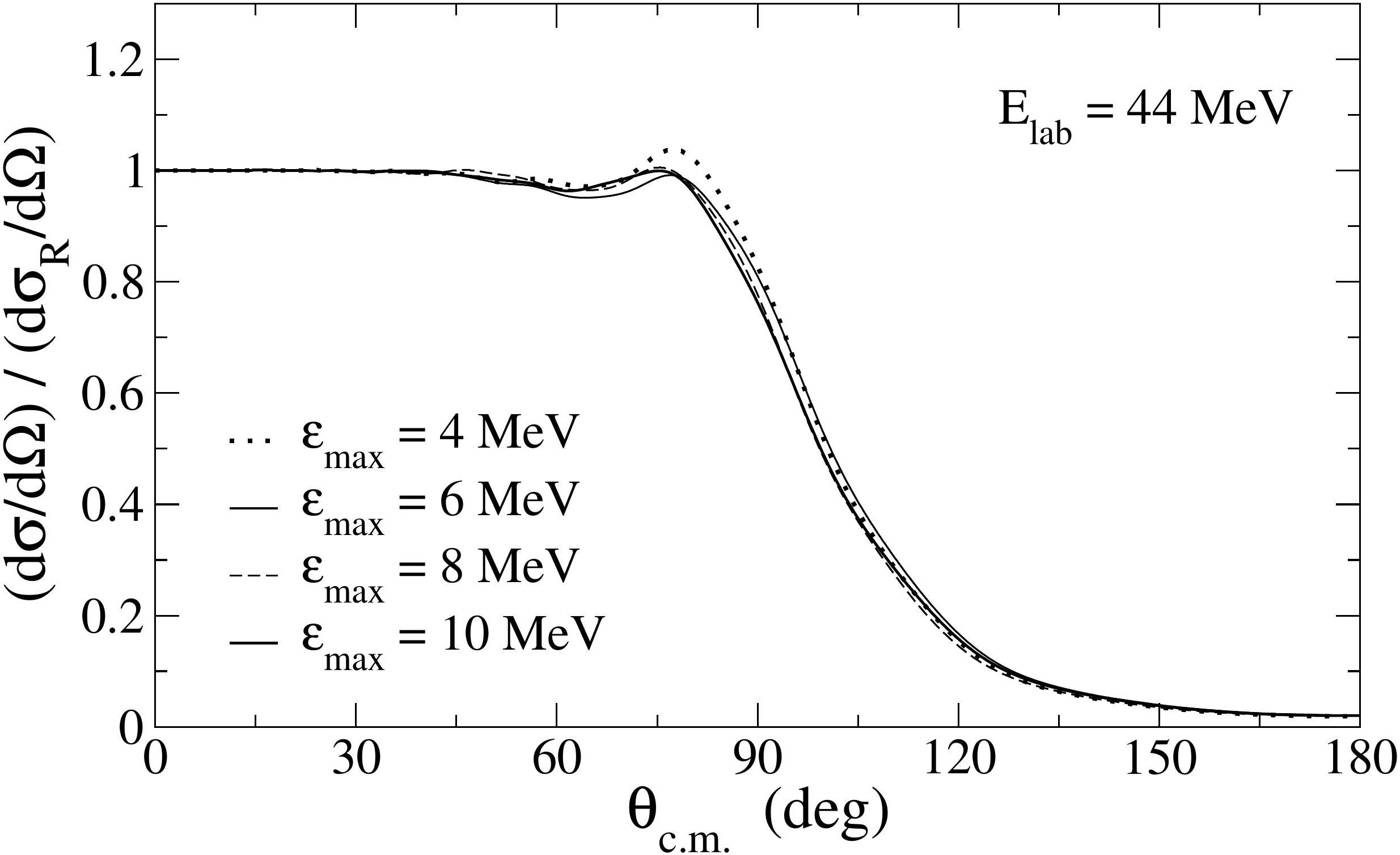}
 \caption{Convergence of the angular distribution of the elastic cross section relative to Rutherford with respect to $\varepsilon_{max}$ for the reaction $^9$Be + $^{208}$Pb at $E_{lab}=44$ MeV. Calculations are performed with $i_{max}=6$.}
 \label{fig:44MeV_emax}
\end{figure}

\begin{figure}
 \includegraphics[width=1\linewidth]{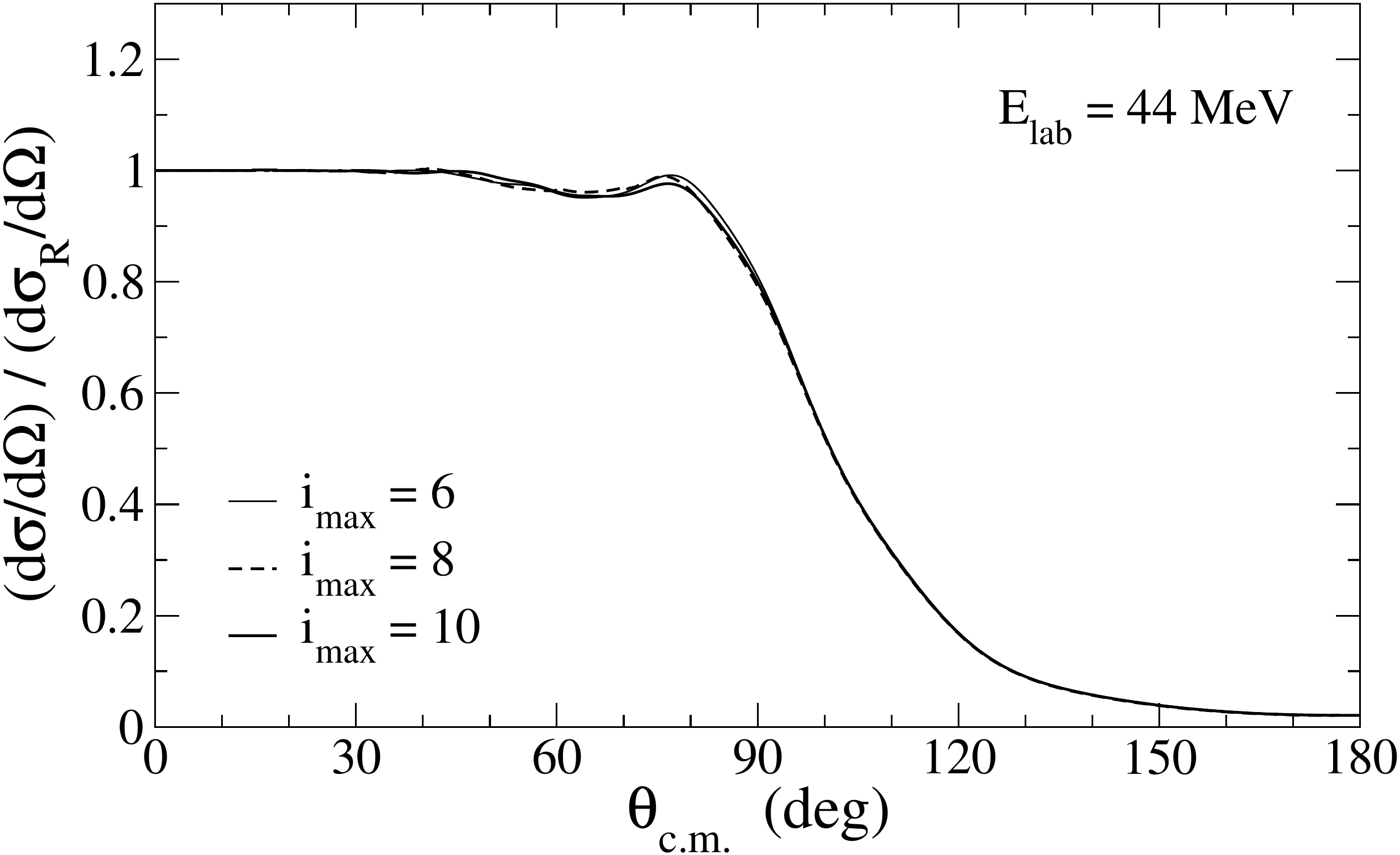}
 \caption{Convergence of the angular distribution of the elastic cross section relative to Rutherford with respect to $i_{max}$ for the reaction $^9$Be + $^{208}$Pb at $E_{lab}=44$ MeV. Calculations are performed with $\varepsilon_{max}=8$ MeV.}
 \label{fig:44MeV_imax}
\end{figure}

Our model space to describe the $^9$Be projectile includes $j^\pi=3/2^{\pm},1/2^{\pm},5/2^{\pm}$ states up to a maximum energy $\varepsilon_{max}$. The states are obtained with a THO basis with maximum 
hypermomentum $K_{max}$ and $i_{max}$ hyperradial excitations in each channel. We first show the convergence of our calculations with respect to the hypermomentum. In Fig.~\ref{fig:60MeV_kmax} we show the elastic cross section at $E_{lab}=60$ MeV
calculated with different values $K_{max}=6,8,10$ for the same value of $i_{max}=6$. In all the cases, we adjust the three-body force in the model Hamiltonian in order to recover the same energy and radius 
of the $3/2^-$ ground state and the same position of different projectile resonances. The calculations with $K_{max}=8$ and 10 are almost identical, confirming the convergence of the results with respect to this parameter. The same 
behavior is observed for the reaction at lower energies. At $E_{lab}=60$ MeV, calculations show a very fast convergence with respect to $i_{max}$ and the cutoff energy $\varepsilon_{max}$. For the reaction around and below the Coulomb barrier, however, a slower convergence is observed. 
To illustrate this point, in Fig.~\ref{fig:44MeV_emax} we 
show the convergence with respect to $\varepsilon_{max}$ of the elastic cross section at $E_{lab} = 44$ MeV, for a fixed value of $i_{max}=6$. 
Calculations with $\varepsilon_{max}=8$ MeV and 10 MeV are 
almost indistinguishable. For calculations with larger $i_{max}$ values, the same behavior is observed. In Fig.~\ref{fig:44MeV_imax} we show the dependence of the calculation at $E_{lab} = 44$ MeV on  the parameter $i_{max}$, with a fixed cutoff energy of 8 MeV. Calculations are 
very close, with small differences only in the angular region between 60 and 90$^\circ$. Working with $i_{max}>10$ is computationally very time-consuming, and we do not expect the results to change significantly. The same features are observed at 
$E_{lab}=38$ MeV. The slower convergence at low energies was already reported for reactions induced by weakly-bound projectiles on heavy targets~\cite{MRoGa08}.

\begin{figure}
 \includegraphics[width=1\linewidth]{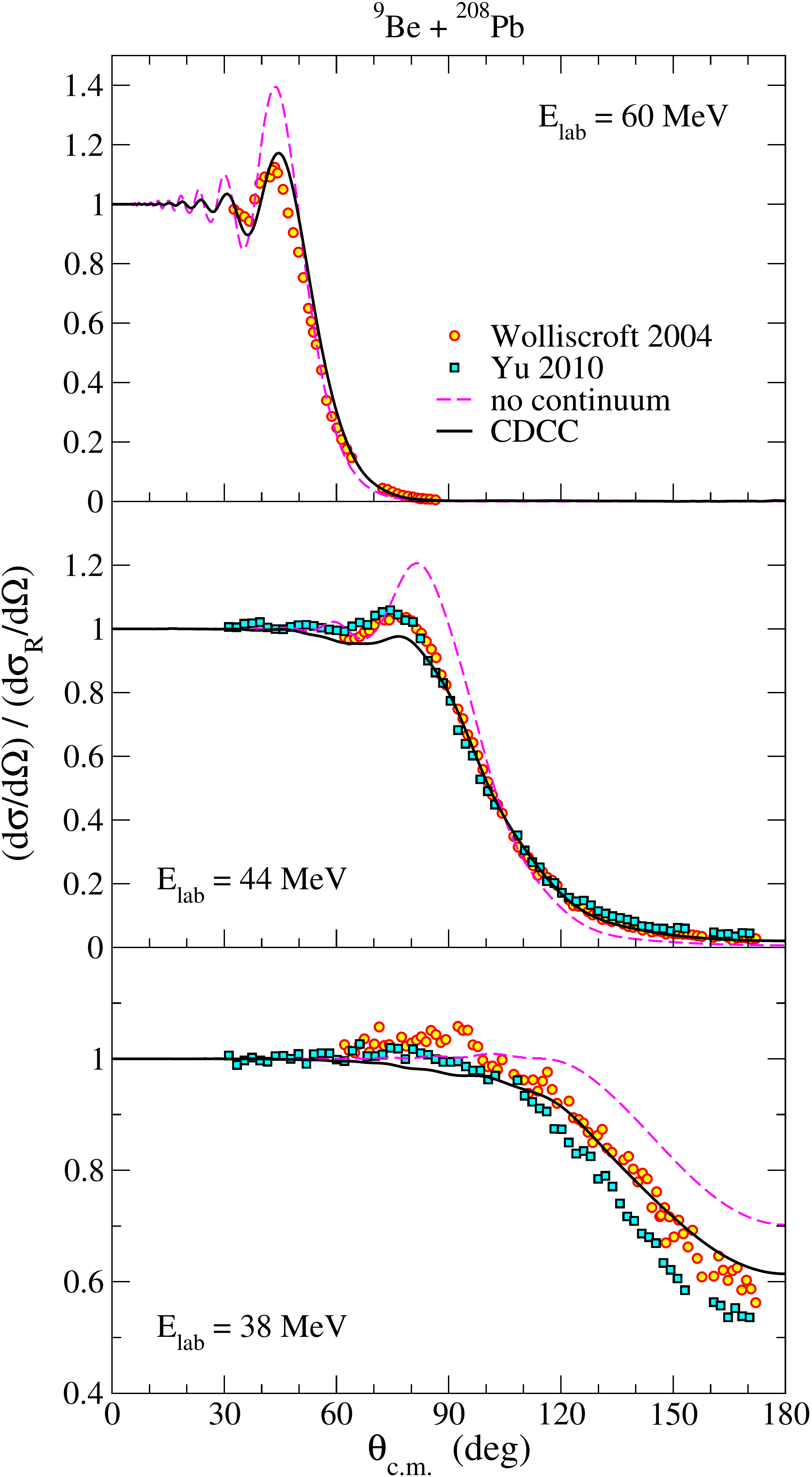}
 \caption{(Color online) Angular distribution of the elastic cross section relative to Rutherford for the reaction $^9$Be + $^{208}$Pb at $E_{lab}=60$, 44 and 38 MeV. Dashed lines correspond to calculations including the ground state only, and solid lines are the full CDCC calculations. The experimental data are shown with circles (Wolliscroft 2004:~\cite{Wolliscroft04}) and squares (Yu 2010:~\cite{Yu10}).}
 \label{fig:all208pb}
\end{figure}

In Fig.~\ref{fig:all208pb} we show our final results at $E_{lab} = 60$, 44 and 38 MeV compared with the experimental data from Refs.~\cite{Wolliscroft04,Yu10}. The error bars are very small and, for clarity, they are not shown. Calculations use $i_{max}=6$, $\varepsilon_{max} = 6$ MeV for the reaction at $E_{lab}=60$ MeV and $i_{max}=10$, $\varepsilon_{max} = 8$ MeV at $E_{lab}=44$ and $38$ MeV. In this figure, dashed lines correspond to calculations including the ground state only, and solid lines are the full CDCC calculations. In all cases, the agreement between our calculations and the data is improved when we include the coupling to breakup channels. The calculations describe reasonably well the experimental data in the complete angular range. Around ($E_{lab} = 44$ MeV) and below (38 MeV) the Coulomb barrier, there is an uncertainty related to the difference between the two data sets. This indicates a possible data normalization problem. At $E_{lab} = 44$ MeV, our calculation underestimate the data between 60 and 90$^\circ$, i.e.~in the nuclear-Coulomb interference region. At $E_{lab} = 38$ MeV, our calculation seems to overestimate the data at backward angles and slightly underestimate the data in the nuclear-Coulomb interference region.

\begin{figure}
 \includegraphics[width=1\linewidth]{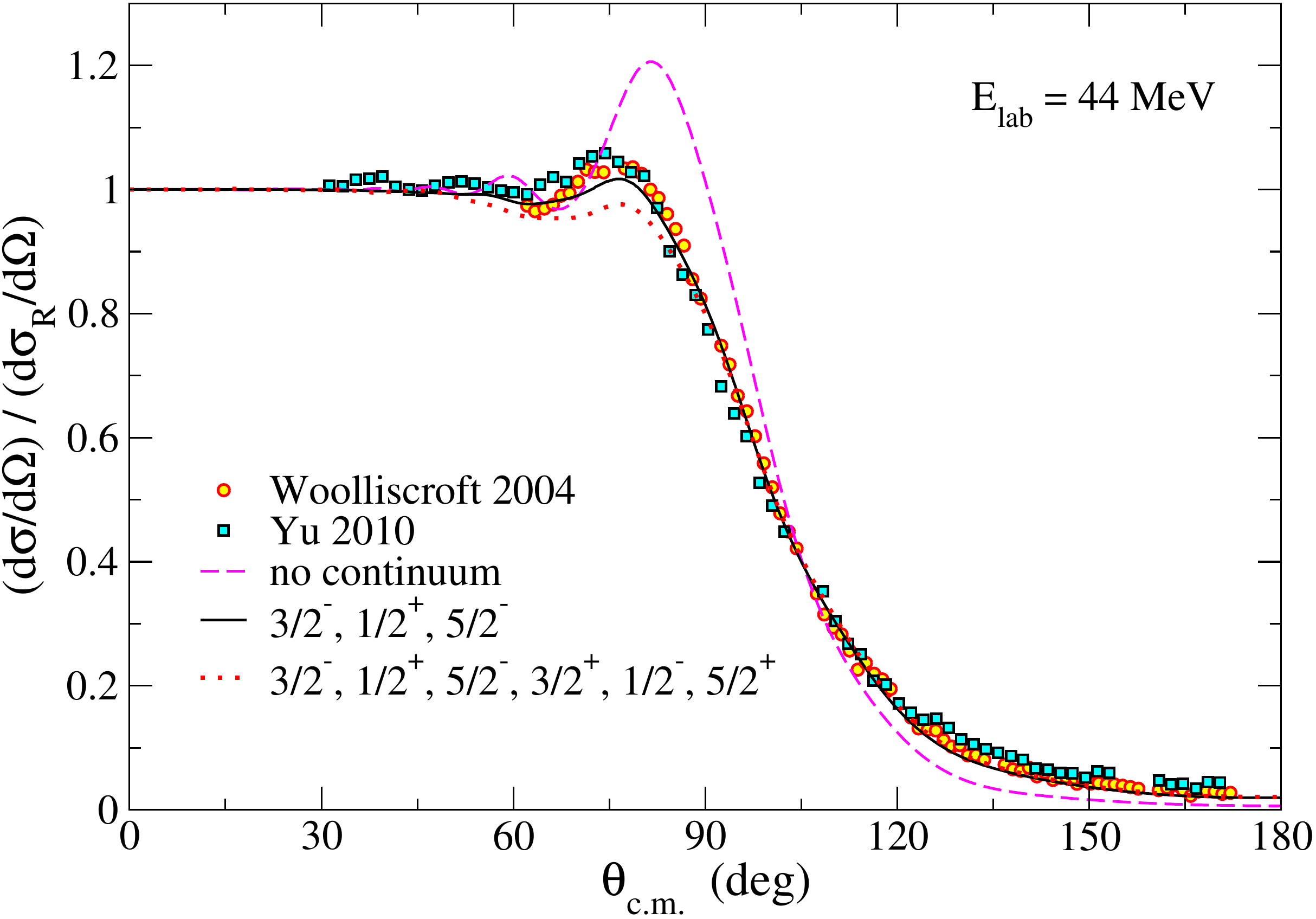}
 \caption{(Color online) Effect of the model space truncation on the elastic cross section relative to Rutherford for the reaction $^9$Be + $^{208}$Pb at $E_{lab}=44$ MeV. Calculations use $i_{max}=10$ and $\varepsilon_{max}=8$ MeV.}
 \label{fig:modelspace}
\end{figure}

It is worth mentioning that these features are not observed in a recent work by Descouvemont \textit{et al.}~\cite{Descouvemont1} using also pseudo-states to describe the continuum. In that work, the Lagrange-mesh basis is used, and different $\alpha$--n, $\alpha$--$\alpha$ potentials are considered. Calculations in~\cite{Descouvemont1} seem to match the experimental data in the nuclear-Coulomb interference region at both $E_{lab} = 44$ and 38 MeV. However, these calculations involve only $j^\pi=3/2^{-},1/2^{+},5/2^{-}$ states. If we use in our calculations just the same $^{9}$Be angular momenta, then we also reproduce the experimental data in the rainbow region. This is shown for clarity in Fig.~\ref{fig:modelspace}. However, in the most recent publication by the same authors~\cite{Descouvemont15}, calculations involve the same model space ($j^\pi=3/2^{\pm},1/2^{\pm},5/2^{\pm}$ states) included in the present work. Results therein are in good agreement with our calculations and show the above-mentioned underestimation of the data in the interference region. Thus, both theoretical approaches are consistent if the same model space is used. A limited model space including only $j^\pi=3/2^-,1/2^+,5/2^-$ states is not sufficient to reach convergence.

The source of the discrepancies, between the converged calculations (Ref.~\cite{Descouvemont15} and the present work) and the experiment in the nuclear-Coulomb interference region, could be due to either the experimental data analysis or the theoretical models used. First, we expect that the scattering of a weakly-bound nucleus such as $^9$Be on a heavy target at energies around and below the Coulomb barrier follows the same behavior reported both, experimentally and theoretically, for other weakly-bound nuclei such as  $^6$He~\cite{Aguilera01,SanBen08}, $^{11}$Li~\cite{Cubero12} and $^{11}$Be~\cite{MGB15}. All these nuclei present a suppression of the rainbow at the interference region when colliding with heavy targets, at energies around and below the Coulomb barrier. This is due to the strong dipolar Coulomb coupling to the continuum states. This suppression is not present in the experimental data on $^9$Be+$^{208}$Pb at $E_{lab} = 44$ MeV and is smaller than the theoretical predictions at $E_{lab} = 38$ MeV.

Concerning the data analysis, a small uncertainty in the angle determination can produce an important deviation on the measured elastic cross section with respect to Rutherford at small angles. This is due to the elastic cross section behavior ($\sin^{-4}(\theta/2)$). In particular, it is necessary to be extremely careful with the beam misalignment. This issue has been addressed in different experimental works (see for example Refs.~\cite{Kakuee03,SanBen08,Cubero12}) and could imply a slope change in the elastic cross section in the rainbow region. On the theoretical side, models describing this kind of reactions depend on several approximations, including the use of optical potentials between the projectile fragments and the target and a truncation of the model space describing the projectile states.
At this point we cannot asses whether the source of the discrepancy arises from experimental problems or theoretical issues. 
However, CDCC calculations agree with the experimental data, elastic and breakup, in the rainbow region for other weakly-bound projectiles under similar scattering conditions (heavy target, energy around the Coulomb barrier). 
This is the case of $^{11}$Li+$^{208}$Pb~\cite{Cubero12,Juanpi13} and $^{11}$Be+$^{197}$Au~\cite{MGB15}, in which the observed  enhancement of the breakup cross section in the nuclear-Coulomb interference region produces a systematic reduction of the elastic cross section at the same angles.
Regardless, differences between the data on $^9$Be+$^{208}$Pb and the four-body CDCC calculations at this region are below 6\%, and the overall agreement is quite good
considering that there is no parameter fitting. This affirms the reliability of a three-body model to describe the structure of $^9$Be.


\begin{figure}
 \includegraphics[width=1\linewidth]{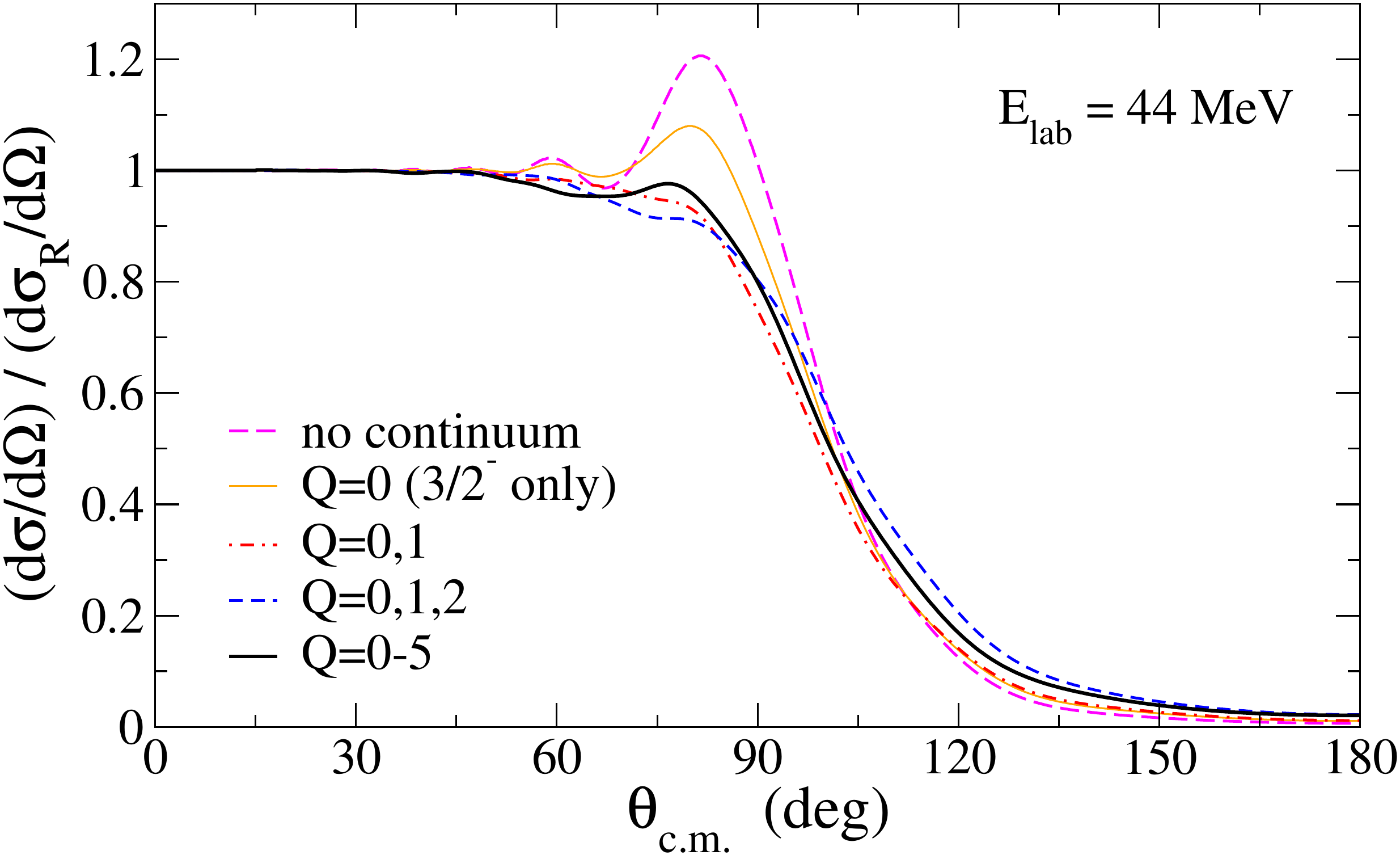}
 \caption{(Color online) Effect of the different coupling multipolarities $Q$ on the elastic cross section relative to Rutherford for the reaction $^9$Be + $^{208}$Pb at $E_{lab}=44$ MeV. Calculations use $i_{max}=10$ and $\varepsilon_{max}=8$ MeV. See the  text for details.}
 \label{fig:44MeV_j}
\end{figure}

In order to study the effect of the $j^\pi$ contributions and coupling multipolarities $Q$ on the results, we show in Fig.~\ref{fig:44MeV_j} different calculations at $E_{lab}=44$ MeV. The monopolar ($Q=0$) contribution allows to connect the 3/2$^-$ ground state to the 3/2$^-$ continuum. Then, the dipolar ($Q=1$) contribution connects the ground state with 1/2$^+$, 3/2$^+$ and 5/2$^+$ states. From them, dipolar and higher order contributions introduce couplings between all $j^\pi$ configurations considered. We see in Fig.~\ref{fig:44MeV_j} that the main contributions to reduce the cross section, the monopole and dipole terms, are of the same order. This result differs from  the case of $^6$He and $^{11}$Li on $^{208}$Pb, where dipolar contributions produce the largest reduction with respect to the calculation without continuum couplings~\cite{MRoGa08,Cubero12}. This dipole effect in halo nuclei, such as  $^6$He and $^{11}$Li,  is due to the deviation of the center of charge with respect to the center of mass because of the presence of two valence particles far away from the charged core ($^4$He or $^9$Li). This produces a strong dipole moment and large $B(E1)$ strengths at low energy. On the contrary, the $^9$Be system is not a halo nucleus and does not present such a large deviation; so dipolar effects, although present, are smaller. This is related to the fact that the sum rule for dipolar transitions is smaller in this case~\cite{JCasal14}. Higher order contributions, specially the quadrupolar terms, produce a correction which improves the description of the experimental data at backward angles.

\begin{figure}[!h]
 \includegraphics[width=1\linewidth]{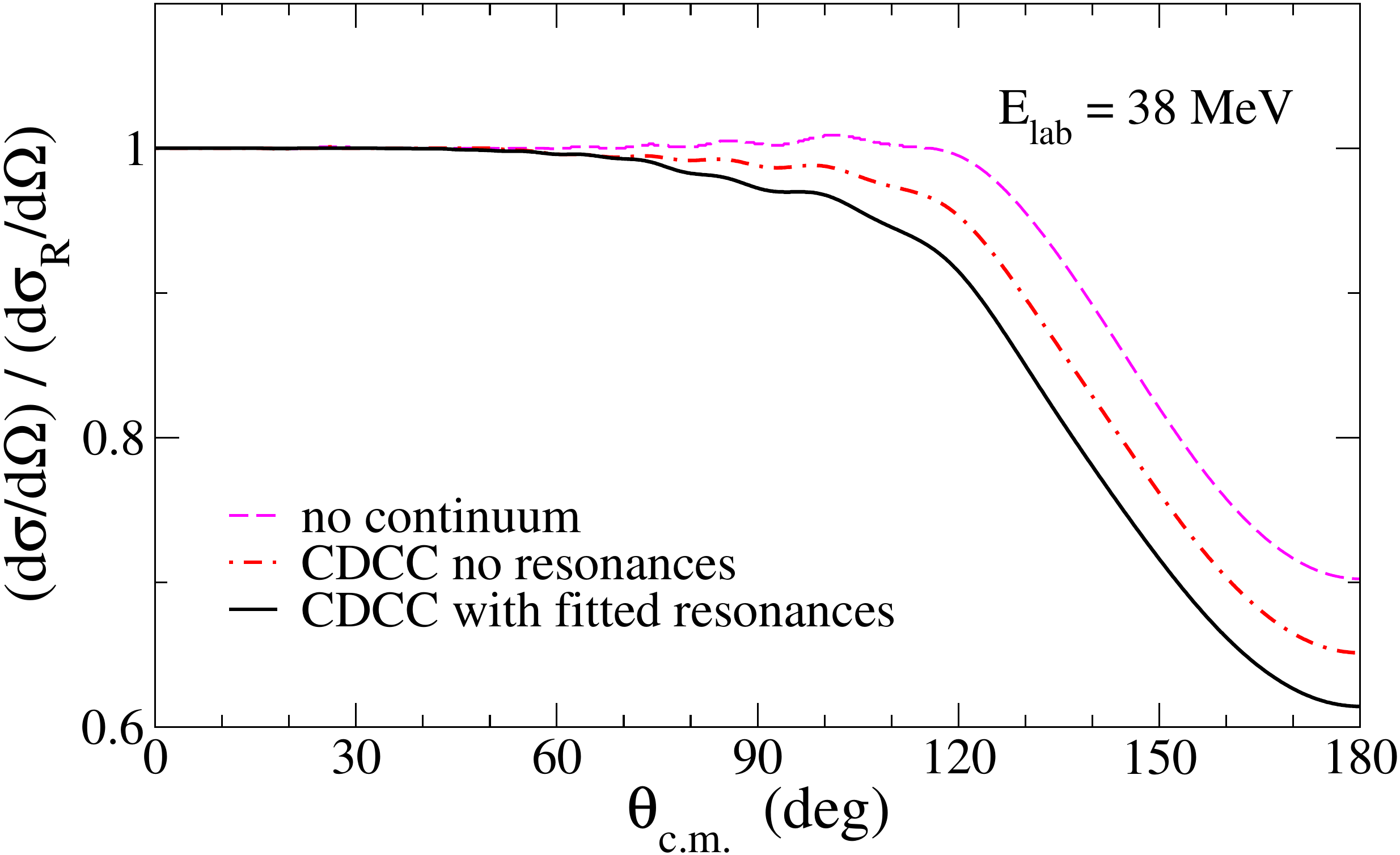}
 \caption{(Color online) Effect of the position of the resonances on the elastic cross section for the reaction $^9$Be + $^{208}$Pb at $E_{lab}=38$ MeV. Calculations use $i_{max}=10$ and $\varepsilon_{max}=8$ MeV.}
 \label{fig:respos}
\end{figure}

We have also studied the effect of the projectile resonances on the elastic cross section. It is known that the low-energy resonances may play an important role in the description of the elastic and breakup processes~\cite{Cubero12}. To illustrate this point, we 
can change the resonance positions by introducing a different three-body strength in Eq.~(\ref{eq:3bforce}) for the 1/2$^{\pm}$, $5/2^{\pm}$ and $3/2^{+}$ states. In Fig.~\ref{fig:respos} we show our calculations ``without'' resonances, i.e.\ including a strong repulsive 
three-body force so that the resonances appear at very high energies and play no role in the CDCC calculations. In that case the cross section exhibits a smaller reduction than the calculation with the resonance positions fitted to the experimental values. Although we cannot separate the resonant and non-resonant parts of the spectrum directly, the calculations clearly show the relevance of the resonance positions in the reaction mechanism.

\subsection{$^9$Be $+$ $^{27}$Al}

\begin{figure}
 \includegraphics[width=1\linewidth]{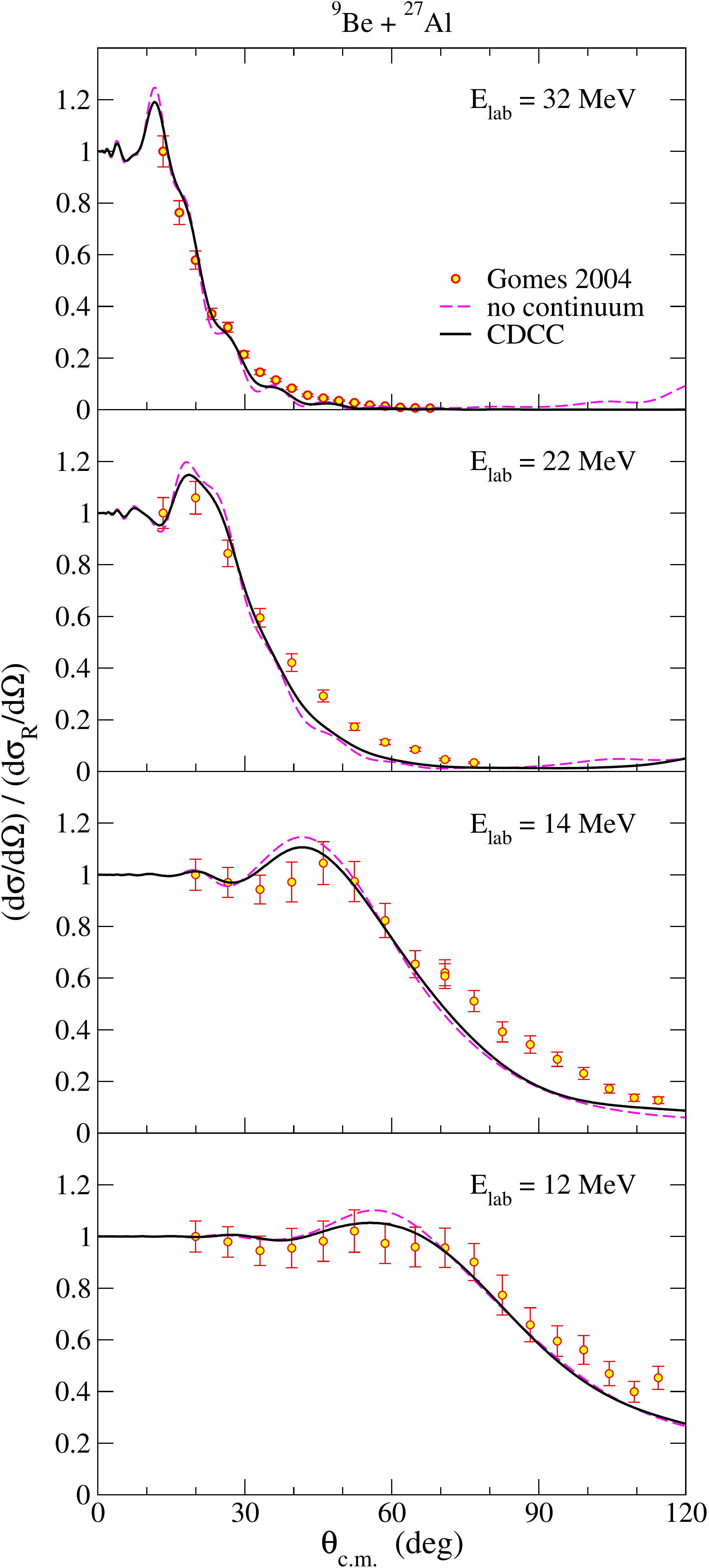}
  \caption{(Color online) Angular distribution of the elastic cross section relative to Rutherford for the reaction $^9$Be + $^{27}$Al at $E_{lab}=32, 22, 14$ and 12 MeV. Calculations use  $i_{max}=6$ and $\varepsilon_{max}=6$ MeV. The experimental data are shown with circles (Gomes 2004:~\cite{Gomes04}).}
 \label{fig:27Al1}
\end{figure}

In order to study the effect of the target mass on the reaction mechanism, we describe in this subsection the scattering of $^9$Be on a lighter  target, $^{27}$Al, using the same formalism. Elastic scattering data are available in the literature for this reaction at near barrier energies~\cite{Gomes04}. In this reference it is shown that the coupling to breakup and/or transfer channels may be relevant even at energies below the Coulomb barrier.

For the coupling potentials, we use again the $n$--target interaction from Ref.~\cite{KD} adjusted for $^{27}$Al. For the $\alpha-^{27}$Al interaction we use the code by S. Kailas~\cite{Kailas}, which provides optical model 
parameters for $\alpha$ particles using the results from Ref.~\cite{Atzrott95}. We have observed that calculations with light targets converge faster with respect to parameters $K_{max}$, $i_{max}$ and $\varepsilon_{max}$, due to the weaker Coulomb interaction. We show our results with $K_{max}=10$, $i_{max}=6$ and $\varepsilon_{max}=6$ MeV, including again $j^\pi=3/2^{\pm},1/2^{\pm},5/2^{\pm}$ 
continuum states. These values ensure convergence at the energies considered, $E_{lab}=12,14,22$ and 32 MeV. In this section we show final results only, although we carried out a convergence analysis similar to the case of the $^{208}$Pb target.

In Fig.~\ref{fig:27Al1} we show the elastic cross section above (32, 22 MeV) and around (14, 12 MeV) the Coulomb barrier. In all the cases, the differences between calculations including only the ground state (dashed lines) 
and full CDCC calculations (solid lines) are significantly smaller with respect to the reaction on $^{208}$Pb. This indicates that breakup effects are less important with light targets and confirms that Coulomb breakup is the 
dominant process at low incident energies. The agreement with the experimental data is reasonable but our calculations underestimate the elastic cross section at backward angles, especially at 14 and 22 MeV. For light targets the Coulomb repulsion is weak, so nuclear effects 
begin to dominate, and the internal structure of the target plays a more important role. This problem may be neglected for heavy targets, but a comprehensive study for light targets is needed. Moreover, the nuclear-dominated region depends on the features of the phenomenological optical potentials between the projectile fragments and the target at the corresponding energy per nucleon. This produces an uncertainty in the nuclear potential that makes difficult, in general, to reproduce with high precision the backward-angle region, where nuclear effects dominate.

\section{Summary and conclusions}\label{sec:summary}

The elastic scattering of $^9$Be on $^{208}$Pb and $^{27}$Al at near Coulomb barrier energies has been described within a four-body CDCC formalism, using the analytical THO method to obtain the internal states of the three-body projectile. In order to get convergence, we have included in the description the $j^\pi=3/2^{\pm}, 1/2^{\pm}$ and $5/2^{\pm}$ states of $^9$Be, paying special attention to the position of the relevant states. 
The agreement with the experimental data is quite reasonable considering that there is no parameter fitting in the CDCC calculations. The convergence with respect to the size of the basis and the maximum excitation energy is slower at energies around and below the Coulomb barrier due to the long-range effects of the Coulomb interaction. 

In the case of the $^{208}$Pb target, the inclusion of continuum couplings in the formalism is essential to describe properly the experimental cross sections. As expected for weakly-bound 
nuclei such as $^9$Be, this effect is important even at energies below the Coulomb barrier.  All the $j^\pi$ contributions included play an important role. It is shown that the scattering process of $^9$Be on $^{208}$Pb at low energies is dominated by the monopolar and dipolar couplings. Dipole contributions are weaker than in halo nuclei, as expected. The position of the lowest resonances affects the cross sections, thus confirming their importance for the reaction mechanism. The discrepancies (below 6\%) between the calculations and the experiment in the nuclear-Coulomb interference region could be due to either the beam misalignment problem or the approximations included in the theoretical formalism. However, the rainbow suppression not shown in the experimental data is expected for weakly-bound nuclei at beam energies around the Coulomb barrier.

In the case of the $^{27}$Al target, continuum couplings produce a significant smaller effect on the cross section. This fact can be explained by taking into account the smaller mass (and charge) of the target, which leads to a smaller Coulomb repulsion and, consequently, reduces the Coulomb breakup. Our calculations underestimate the elastic cross-section at backward angles, especially at 14 and 22 MeV, where nuclear effects dominate. In this region, the internal structure of the target may play a significant role, and the calculations depend on the features of the phenomenological optical potentials between the projectile fragments and the target. A further analysis of 
these effects is desirable.  

The results we have obtained encourage the application of the analytical THO method within a four-body CDCC framework to reactions induced by three-body projectiles with more than one charged particle and supports the reliability of our three-body model to describe the structure of $^9$Be.

\begin{acknowledgments}

Authors are grateful to  P. Descouvemont and J. Gómez-Camacho for useful discussions and suggestions. 
This work has been partially supported by the Spanish Ministerio de Econom\'{\i}a y Competitividad under Projects FIS2013-41994-P, FIS2011-28738-c02-01, FIS2014-53448-c2-1-P and FIS2014-51941-P and by Junta de Andaluc\'{\i}a under group number FQM-160 and Project P11-FQM-7632.
J. Casal acknowledges a FPU research grant from the Ministerio de Educaci\'on, Cultura y Deporte, AP2010-3124. M. Rodríguez-Gallardo acknowledges a postdoctoral contract by the V Plan Propio de Investigaci\'on of the Universidad de Sevilla.

\end{acknowledgments}


\bibliography{./bibfile}

\end{document}